\documentclass[12pt]{JHEP3} 

\usepackage{amsmath,amssymb}

\renewcommand{\v}{{\varphi}}
\renewcommand{\o}{{\omega}}

\newcommand{\be}{\begin{equation}}
\newcommand{\ee}{\end{equation}}
\newcommand{\bea}{\begin{eqnarray}}
\newcommand{\eea}{\end{eqnarray}}

\newcommand{\pint}{\makebox[0pt][l]{\hspace{2.4pt}$-$}\int}

\newcommand{\atopfrac}[2]{\genfrac{}{}{0pt}{}{#1}{#2}}
\newcommand{\sfrac}[2]{{\textstyle\frac{#1}{#2}}}
\newcommand{\half}{\sfrac{1}{2}}

\newcommand{\Q}{\mathbf{Q}}
\newcommand{\q}{\mathbf{q}}
\newcommand{\contourgauge}{\mathbf{C}}

\newcommand{\alg}[1]{\mathfrak{#1}}

\newcommand{\su}{\alg{su}}

\title{
Bethe Ansatz for Quantum Strings
}

\author{Gleb Arutyunov$^{1,\dagger}$, Sergey Frolov$^{2,1,\dagger}$ 
and Matthias Staudacher$^{1}$ \\
\vskip 0.5cm
$^{1}$Max-Planck-Institut f\"ur Gravitationsphysik,
Albert-Einstein-Institut\\
Am M\"uhlenberg 1, D-14476 Potsdam, Germany\\
\vskip 0.1cm
$^{2}$Department of Applied Mathematics, \\
SUNY Institute of Technology,\\
P.O. Box 3050, Utica, NY 13504-3050\\

\vskip 0.1cm
Email: \email{agleb@aei.mpg.de; frolovs@sunyit.edu; matthias@aei.mpg.de}   }

\preprint{\hepth{0406256}\\
AEI-2004-046}

\abstract{We propose Bethe equations for the diagonalization of 
the Hamiltonian of quantum strings on $AdS_5\times S^5$ at large
string tension and restricted to certain large charge states 
from a closed $\su(2)$ subsector. The ansatz differs from the 
recently proposed all-loop gauge theory asymptotic Bethe ansatz by additional 
factorized scattering terms for the local excitations. We also
show that our ansatz quantitatively reproduces {\it everything} that 
is currently known about the string spectrum of these states.
Firstly, by construction, we recover the integral Bethe equations 
describing semiclassical spinning strings.
Secondly, we explain how to derive the 1/J energy
corrections of $M$-impurity BMN states, 
provide explicit, general formulae
for both distinct and confluent mode numbers,
and compare to asymptotic gauge theory.
In the special cases $M=2,3$
we reproduce the results of direct quantization of Callan et al. 
Lastly, at large string tension and relatively small charge we
recover the famous $2\sqrt[4]{n^2\lambda}$ asymptotics
of massive string modes at level $n$.
Remarkably, this behavior is entirely determined by the novel
scattering terms. This is qualitatively consistent with the
conjecture that these terms occur due to wrapping effects in
gauge theory. Our finding does {\it not} in itself cure the
disagreements between gauge and string theory, but leads us
to speculate about the structure of an interpolating Bethe
ansatz for the AdS/CFT system at finite coupling and charge.

\vskip 0.5cm
$^{\dagger}$ Also at Steklov Mathematical Institute, Moscow.

}


\begin{document}

\section{Introduction}
Recently remarkable progress in our understanding of the gauge/string 
duality \cite{M} was achieved.  
It was initiated by  Berenstein, Maldacena and Nastase (BMN) \cite{BMN},  
who proposed a way to match energies of certain string states 
with perturbative scaling dimensions of dual SYM operators.

Later on it was found \cite{GKP2,FT} that there exists an even larger 
sector of string states on $AdS_5\times S^5$ which {\it a priori} permits a
direct comparison with perturbative gauge theory.
Indeed a certain region of 
the quantum string spectrum can be well approximated 
by classical string solutions describing highly energetic strings. 
In most cases, however, comparison of the energy of such a string with a
perturbative scaling dimension of the dual SYM operator is impossible
since the string energy generically turns out to be non-analytic in the
't Hooft coupling constant $\lambda$.
Fortunately, there exist solitonic solutions corresponding 
to classical multi-spin strings rapidly rotating in $S^5$ 
whose energies admit an expansion
in integer powers of the effective coupling constant
$\lambda/L^2$, where $L$ is the large, total spin on $S^5$. 
This offers the opportunity to directly compare the energies of 
spinning strings with {\it perturbative} scaling dimensions of gauge 
theory operators. Let us stress, however, that the comparison
is {\it not} guaranteed to be successful, as the coupling
$\lambda$ is nevertheless large in string theory, while it is
small in perturbative gauge theory.

The simplest configurations are folded and circular rigid strings 
and the corresponding solutions 
can be obtained from an integrable system of the Neumann type \cite{AFRT}
which is a finite-dimensional reduction of the
classical string sigma-model on $AdS_5\times S^5$.

In a parallel development, related recent advances in gauge theory
came to be after the important observation that planar, 
conformal ${\cal N}=4$ SYM is an integrable system 
in the one-loop approximation \cite{MZ,BS} and, very likely, at higher 
loops as well \cite{BKS}. This extends and generalizes in a beautiful way 
the previously observed integrable structures
of one-loop QCD \cite{BDM}, which also happens to be conformal.
Integrability is indispensable for resolving the 
complicated mixing problem in order to find the anomalous dimensions 
of conformal operators. The main new tool it provides is the 
{\it Bethe ansatz}, which reduces the spectral problem to the 
solution of a system of finitely many non-linear equations.
For a closed $\su(2)$ subsector of operators the one-loop
Bethe ansatz of \cite{MZ} was extended to three loops in
\cite{SS}, using the three-loop integrable structures of
\cite{BKS}. Very recently an all-loop ansatz (under certain
restrictions to be discussed below) was proposed in \cite{BDS}. 

The Bethe ansatz allowed one to perform very non-trivial comparisons 
of gauge and string theory predictions as first shown in \cite{BFMSTZ} 
in various special cases such as the above folded and circular rigid 
strings. Not only the energies but also the eigenvalues of higher
charges agree at leading one-loop order \cite{AS}, indicating
that the integrable structures on both sides of the correspondence
are closely related. Furthermore, it proved possible to exploit
the classical integrability of the string sigma model and to derive
a Bethe equation describing its semi-classical solutions 
for the simplest case of strings moving in $R\times S^3$ in
full generality \cite{KMMZ}. This led to perfect agreement between 
string and gauge theory structures up to two-loop order of
perturbation theory. Further important work on the ``spinning limit''
of AdS/CFT was performed in \cite{Upp,Kruc,Kop}, and 
other relevant aspects of the gauge/string duality 
have been studied in \cite{Rus}.

Starting at three-loop order this matching pattern 
breaks down both for spinning strings \cite{SS,AS} and strings
in the near plane-wave background \cite{Call1,Call2}. 
One possible explanation was suggested in \cite{SS}, and
refined in \cite{BDS}. Even though the effective coupling constant 
$\lambda/L^2$ is small
the 't Hooft coupling $\lambda$ must be large for semiclassical 
spinning strings, as well as for strings in the near pp-wave
background. Thus, a fully reliable comparison with gauge theory 
would require the complete summation of the
gauge theoretic perturbative expansion, 
$i.e.$ the knowledge of the spectrum of the all-loop dilatation operator. 
In the spin chain picture
the dilatation operator is identified with the Hamiltonian of a 
long-range spin chain. 
In particular, in the closed $\su(2)$ subsector 
the three-loop  dilatation operator 
 coincides \cite{SS} with the Hamiltonian of the integrable Inozemtsev spin
chain \cite{Inoz}. At four loops the Inozemtsev chain violates 
BMN scaling, and, therefore, precludes
perturbative comparison with string theory. Quite remarkably, another 
long-range spin chain exists which is different from the one of
Inozemtsev starting at four loops, does {\it not} violate BMN scaling, 
and leads to the string-theoretic BMN energy formula \cite{BDS}. 
Even though the explicit form of the
all-loop Hamiltonian of this novel spin chain is currently unknown, 
the associated Bethe equations as well as 
the expressions for the eigenvalues of all higher conserved charges
have been proposed in \cite{BDS}. 

One intriguing result of \cite{BDS}, which will play a key role in
this paper, was that the functional dependence of
the individual charges of the elementary excitations (termed charge 
densities in \cite{BDS}) on the excitation momenta {\it agrees} 
in the thermodynamic limit to all loop orders when one compares 
the (conjectured) gauge and (known) string theory expressions
in the $\su(2)$ subsector. What is different 
is the {\it distribution} of excitation momenta. This may be traced back 
to the fact that the respective Bethe equations, whose solution
precisely fixes the excitation momenta, differ by a certain term 
starting at three loop order.

As was discussed in \cite{BDS} this difference might have two possible
explanations. The first, somewhat disappointing one would be that
the AdS/CFT correspondence is only approximate rather than exact.
A second, more exciting one would be that non-perturbative effects
have to be taken into account. In particular it was argued in 
\cite{BDS} that these might arise after the inclusion of so-called 
{\it wrapping interactions} into the gauge theory computations. 
And indeed, the currently proposed Bethe ansatz of \cite{BDS} is 
asymptotic in the sense that it correctly yields the 
gauge theory scaling dimensions only as long as the range of the
spin chain interaction does not exceed the length of the spin chain.  
Unfortunately it is presently quite unclear how to properly account 
for the wrapping interactions on the gauge side.

The long-range Bethe equations of \cite{BDS} and
the classical Bethe equations of \cite{KMMZ} conceptually differ
in another important respect. The former are {\it discrete} equations 
applicable to a finite number of excitations (magnons) and
at finite length (modulo the just explained restriction
of asymptoticity). The latter are {\it continuum} (``thermodynamic'')
equations which assume a macroscopically large length and a large
number of excitations. It is very natural to ask for a discrete
version of the string Bethe equations. These should then properly
describe the {\it quantum} corrections of the string sigma model.
If we assume that integrability survives quantization on the
string side these discrete equations should certainly exist.  
In this paper we will take inspiration from the spin chain
equations to propose precisely such a set of
equations engineered to account for the leading quantum
corrections for strings on $AdS_5\times S^5$.


Our discretization reduces, by construction, in the thermodynamic limit to
the integral Bethe equations describing classical spinning strings 
\cite{KMMZ,BDS}.
We then use these equations to compute the leading finite length correction 
to the two- and three-impurity BMN operators and find remarkable agreement 
with the results of \cite{Call1} and \cite{Call2}. We are also
able to treat the multi-impurity problem in generality, which
seems to be hardly possible in the framework of \cite{Call1,Call2}.
In fact, our Bethe equations diagonalize the quantum 
string Hamiltonian to leading order in 
$1/J$. We then turn to the study of the strong coupling limit. 
We use our equations to show that conformal dimensions of gauge theory 
operators dual to massive string modes at level $n$ exhibit the famous 
$2\sqrt[4]{n^2\lambda}$ asymptotics. 
Remarkably, this behavior is completely due to the terms absent in the
asymptotic gauge ansatz. While this fact of course does not
prove the wrapping scenario, it is certainly fully consistent with it.
In summary, the Bethe equations we propose are compatible with our 
current understanding of the spectrum of string theory on $AdS_5\times S^5$.
  
Our proposal is not yet complete as it is likely to need further
refinement in order to include higher quantum string effects.
However, we do speculate on a possible general form of the
full quantum Bethe ansatz at the end of this paper.  
It is based on the idea that the correct ansatz should,
if AdS/CFT is correct, interpolate between the string ansatz at
large string tension $\sqrt{\lambda}$ and the gauge ansatz at small
't~Hooft coupling $\lambda$.
It would be important to find further evidence in its favor.

\section{Bethe Ansatz for Quantum Strings}
\subsection{The Proposal}
As was discussed in the introduction 
the classical sigma-model describing strings on $AdS_5\times S^5$ 
is an integrable system.
One may hope that integrability is also preserved on the 
quantum level. In the $\su(2)$ subsector of string states with two 
spins, $M$ and $L-M$, the spectrum is encoded 
in a set of integral equations of Bethe type \cite{KMMZ}. Thus, 
it is natural to expect that the quantum spectrum should be also 
described by a system of Bethe equations. These
equations should diagonalize the quantum string Hamiltonian.
Unfortunately the exact quantization of string theory is beyond our 
present reach.
However, one can try to make an educated guess for such a system of equations, 
using our knowledge of the Bethe equations describing 
classical spinning strings and the asymptotic Bethe equations conjectured for 
the perturbative gauge theory.

\medskip 

The Bethe equations we propose for describing the leading quantum effects for
strings in the $\su(2)$ sector have the following form
\bea
\exp(iLp_k)
=
\prod_{\textstyle\atopfrac{j=1}{j\neq k}}^M S(p_k,p_j)\, , ~~~~~\sum_{k=1}^Mp_k=0 \, .
\label{BAB}
\eea
where the matrix $S(p_k,p_j)$ describing the pairwise scattering
of local excitations with momenta $p_k$ is given by 
\bea
\label{SM}
S(p_k,p_j)&=&
 \frac{\varphi(p_k)-\varphi(p_j)+i}{\varphi(p_k)-\varphi(p_j)-i}\times \\
&\times &
\exp\Big(2i\sum_{r=0}^{\infty}\Big({\frac{g^2}{2}}\Big)^{r+2}
\big(\q_{r+2}(p_k)\q_{r+3}(p_j)-\q_{r+3}(p_k)\q_{r+2}(p_j)\big)
\Big)
\,
 .
\nonumber
\eea
Here $M$ is the total number of local excitations and 
the phase function $\v(p)$ is 
\bea
\label{pf} \varphi(p)=\half\cot(\half p )\sqrt{1+8g^2\sin^2(\half
p)} \,\, , \eea where the coupling constant $g$ is related to the
't Hooft coupling $\lambda$ as
$$
g^2 = {\frac{\lambda}{8\pi^2}}\ .
$$
The functions $\q_r(p)$ (charges) are given by
\bea \label{qr}
\q_r(p)=g^{-r+1}\frac{2\sin(\frac{1}{2}(r-1)p)}{r-1}
\left(\frac{\sqrt{1+8g^2\sin^2(\frac{1}{2}p)}\, -\, 1
}{2g\sin(\frac{1}{2}p)} \right)^{r-1} \, .\eea 
In particular,  the first charge
$\q_1(p)$ is the momentum $p$ and the second one $\q_2(p)$
is the energy of a single excitation
\bea\label{q2}
\q_2(p)=\frac{1}{g^2}\left(\sqrt{1+8g^2\sin^2(\half p)}\, -\, 1
\right)\, . \eea
In the sequel we will need the total charge of the $M$ excitations
given by the sum of individual charges 
\bea \label{Qr}
\Q_r=\sum_{k=1}^M\q_r(p_k) \,  . 
\eea
We expect this model to reproduce the quantum spectrum of string states 
with $L$ and $g$ large. The energy of string states in the global 
AdS coordinates is proposed to be 
\bea \label{Ener} {\bf E}(g)=L+g^2 \Q_2 \, . \eea

The phase function $\v(p)$ and the charges $\q_r(p)$ first appeared in \cite{BDS}
where an all loop Bethe ansatz diagonalizing the dilatation operator 
in the $\su(2)$ subsector of ${\cal N}=4$ SYM theory in the asymptotic limit 
$L\to\infty$ was proposed. In fact our equations (\ref{BAB}) differ
from the asymptotic Bethe ansatz of \cite{BDS} by the exponential term in the 
scattering matrix $S(p_k,p_j)$. Note that in the spin chain picture of \cite{BDS}
$L$ is identified with the length of the long-range spin chain, while $M$
is the number of magnons. The dilatation operator is related to the second charge 
$\Q_2$ in the same way as in eq.(\ref{Ener})
\bea \label{Dim} {\bf D}(g)=L+g^2 \Q_2 \, .\eea

Our proposal for the quantum Bethe ansatz is inspired by the observation of 
\cite{BDS} that in the thermodynamic limit 
elementary excitations of gauge and string theory coincide. This motivates us
to modify only the scattering matrix but to  
keep the same  charges and phase function as in the gauge theory asymptotic Bethe 
ansatz.  According to the 
AdS/CFT correspondence the string energy is equal to the scaling dimension 
of the dual gauge theory operator. In spite of the fact that we change the Bethe 
equations we still use the same relation between the energy/scaling dimension 
and the second conserved charge as in the asymptotic Bethe ansatz, {\it i.e.},
as in perturbative gauge theory.

\subsection{Heuristic Derivation}
Here we present heuristic arguments motivating our proposal.
We start from the integral Bethe equation which describes the 
classical spinning strings in the $\su(2)$ subsector \cite{KMMZ}.
As was shown in \cite{BDS} the integral Bethe equations of \cite{KMMZ}
can be written as follows
\bea
\label{inBA}
&&\pint_{\contourgauge}
\frac{d\varphi'\rho_{\mbox{\scriptsize{s}}}(\v')}{\varphi-\varphi'}
=\frac{1}{2}\frac{1}{\sqrt{\varphi^2-4\o^2}}+\pi n_{\nu}+2\o^2\int_\contourgauge
\frac{d\varphi'\rho_{\mbox{\scriptsize{s}}}(\v')}{\sqrt{\varphi^2-4\o^2}
\sqrt{\varphi'^2-4\o^2}} \times~~~~~~~~\\
\nonumber
&&~~~~~~~~~~~~~~~~~~~~~~~~~~~~~~~~~~\times
\frac{\v-\sqrt{\varphi^2-4\o^2}-\v'+\sqrt{\varphi'^2-4\o^2}}
{(\v+\sqrt{\varphi^2-4\o^2})(\v'+\sqrt{\varphi'^2-4\o^2})-4\o^2}\, ,
\eea
where $n_{\nu}$ are winding numbers, and
$\rho_{\mbox{\scriptsize{s}}}(\v)$ is the spectral density of a
finite-gap solution of the string sigma-model. 
The spectral density has a support on a union 
$\contourgauge$ of smooth contours in the complex $\v(p)$-plane and normalized as
\bea
\label{normal}
\int_{\contourgauge} d\v\ \rho_{\mbox{\scriptsize{s}}}(\v ) 
=\frac{M}{L}=\alpha \, .
\eea 
In eq.(\ref{inBA}) the effective coupling constant $\o$ is defined as
$$
\o^2 = {\frac{g^2}{2L^2}}\ .
$$
Recall that the last term on the r.h.s.~of eq.\eqref{inBA} is
absent in the conjectured asymptotic gauge theory ansatz.
Expanding the denominators of the last term in 
eq.(\ref{inBA}) in a geometric series, we can rewrite this equation
in the form
\bea
\nonumber
&&\pint_{\contourgauge}
\frac{d\varphi'\rho_{\mbox{\scriptsize{s}}}(\v')}{\varphi-\varphi'}
=\frac{1}{2}
\frac{1}{\sqrt{\varphi^2-4\o^2}}+\pi n_{\nu}+
\sum_{r=0}^\infty \o^{2r+4}\int_\contourgauge
\frac{d\varphi'\rho_{\mbox{\scriptsize{s}}}(\v')}{\sqrt{\varphi^2-4\o^2}
\sqrt{\varphi'^2-4\o^2}} \times\\
\label{CSBE}
&&
~~~~~~~~~~~~~~~~\times \left( \frac{1}{ \left(\frac{1}{2}\v+\frac{1}{2}
\sqrt{\varphi^2-4\o^2}\right)^{r+2}
\left(\frac{1}{2}\v'+\frac{1}{2}\sqrt{\varphi'^2-4\o^2}\right)^{r+1}}\ -\right. \\
\nonumber
&&
~~~~~~~~~~~~~~~~~~~~~~~~~~~~\left. \frac{1}{ \left(\frac{1}{2}\v+\frac{1}{2}
\sqrt{\varphi^2-4\o^2}\right)^{r+1}
\left(\frac{1}{2}\v'+\frac{1}{2}\sqrt{\varphi'^2-4\o^2}\right)^{r+2}} \right)\, .
\eea
According to \cite{BDS} the 
commuting charge densities $\q_r(\phi)$ of the string 
sigma-model have the form
\bea
\label{qrt}
\q_r (\v) = \frac{1}{\sqrt{\varphi^2-4\o^2}}\frac{1}
{\left(\frac{1}{2}\v+\frac{1}{2}\sqrt{\varphi^2-4\o^2}\right)^{r-1}}\, ,
\eea
where in particular $p(\v)=\q_1(\v)$. In terms of the spectral density 
the total commuting charges take the form  
\bea
\label{Qrt}
\Q_r = \int_\contourgauge\ d\v\ \rho_{\mbox{\scriptsize{s}}} (\v )\ \q_r(\v)\, .
\eea
Taking into account the expression for the charges we see that eqs.(\ref{CSBE})
can be cast into the following form 
 \bea
\label{our}
\pint_{\contourgauge}\frac{d\varphi'\rho_{\mbox{\scriptsize{s}}}(\v')}{\varphi-\varphi'}
=\pi n_{\nu}+\frac{1}{2}p(\v)+\sum_{r=0}^\infty \o^{2r+4}\left(
\q_{r+3}(\v)\Q_{r+2} -\q_{r+2}(\v)\Q_{r+3}
 \right)\, .
\eea
Now we note that this integral equation arises in the thermodynamic limit from 
the following discrete set of equations
 \bea
&&\exp\Big(iLp_k+2i\sum_{r=0}^{\infty}\Big(\frac{g^2}{2}\Big)^{r+2}
(\q_{r+3}(p_k)\Q_{r+2}-\q_{r+2}(p_k)\Q_{r+3}
)\Big)=~~~~~~~~~~~~~~~~~~~~~ \label{BAI}
\\
&&~~~~~~~~~~~~~~~~~~~~~~~~~~~~~~~~~~~~~~~~~~~~~~~~~~~~~~~~~~~~~~
=
\prod_{\textstyle\atopfrac{j=1}{j\neq k}}^M
 \frac{\varphi(p_k)-\varphi(p_j)+i}{\varphi(p_k)-\varphi(p_j)-i}\,
 ,
\nonumber \eea
where the charges $\q_r(p)$, $\Q_r$ and the phase function $\v(p)$
are given by eqs.(\ref{qr}), (\ref{Qr}) and (\ref{pf}). 
Indeed, in the thermodynamic limit
$M$ and $L$ go to infinity with the ratio $\alpha=\frac{M}{L}$ held fixed.
In this limit momentum $p_k$, 
charges $\q_r(p_k)$ and $\Q_r$ scale as $p_k\to p_k/L$, $\q_r(p_k)\to L^{-r}\q_r(p_k)$
and  $\Q_r\to L^{-r+1}\Q_r$ respectively. The rescaled 
phase function $\v\to \v/L$ acquires the form
\bea
\v (p) = \frac{1}{p}\sqrt{1+4\o^2 p^2} \, .
\eea    
To compare with the integral Bethe equation (\ref{our})
it is useful to express $p$ and all other charges as functions of $\v$. 
Then one can see that the rescaled charges $\q_r(\v)$ coincide 
with (\ref{qrt}). Finally to obtain 
the integral Bethe equations we introduce the 
distribution density
\bea
\label{den}
\rho(\v)=\frac{1}{L}\sum_{k=1}^M\delta(\v-\v(p_k))\, ,\qquad
\int_{\contourgauge} d\v\ \rho (\v ) =\frac{M}{L}=\alpha \, .
\eea 
Now taking the thermodynamic limit of eqs.(\ref{BAI}) and identifying $\rho$
with $\rho_{\mbox{\scriptsize{s}}}$ we obtain the integral equation (\ref{our}).

Due to the presence of the total charges
it may be not clear why (\ref{BAI}) is compatible with the
principle of factorized scattering.   
However, taking into account formula (\ref{Qr}) expressing the total charge 
$\Q_r$ as the sum of $\q_r(p_k)$ we see that eq.(\ref{BAI})
is equivalent to eqs.(\ref{BAB}), (\ref{SM}) if we bring the charge-dependent term 
on the r.h.s. In this form the factorization property
becomes transparent.
Indeed, the phase shift acquired by an individual excitation
traveling around a circle of length $L$ is equal to the sum of pairwise  
phase shifts which arise due to its elastic scattering with the
other $M-1$ excitations. 

Concluding this section let us note that in the BMN limit, where the number $M$ 
of elementary excitations is kept finite,  
both $\q_r(p_k)$  and $\Q_r$ scale as  $\q_r(p_k)\to L^{-r}\q_r(p_k)$
and  $\Q_r\to L^{-r}\Q_r$ respectively. Therefore, the additional exponential 
term in the scattering matrix (\ref{SM})
appears to be $1/L$ suppressed and drops out in the strict 
$L\to \infty$ limit. Therefore, in the BMN limit the resulting quantum Bethe equations 
coincide with the ones of the  
asymptotic Bethe ansatz and lead to the BMN energy formula (see \cite{BDS}). 
However, the exponential term becomes relevant in considering 
$1/L$ corrections, {\it i.e.} in the near-BMN limit. We will now
turn our attention to this limit.

\section{Near-BMN Limit}

\subsection{Separated Mode Numbers}

The direct quantization of string theory in the BMN limit \cite{BMN} is 
feasible since the complicated curved background $AdS_5 \times S^5$ may
be replaced by the much simpler pp-wave background. In lightcone 
gauge it allows for an exact free field quantization, 
as first shown by Metsaev and Tseytlin \cite{METS}. The string
predictions for the $\su(2)$ sector with $M$ impurities\footnote{
In the condensed matter literature the elementary excitations 
of a ferromagnetic spin chain are certainly never denoted as
``impurities'' but rather as magnons. On the string side
the nature of the elementary excitations at finite $L$ and $M$
is not known, but presumably related to some kind of ``string bits''.} 
are immediately reproduced, by construction, on the gauge side 
if one uses the conjectured asymptotic long range 
spin chain Bethe ansatz of \cite{BDS}. One keeps $M=2,3, \dots$ 
finite, puts $L=J+M \sim J \rightarrow \infty$,
and holds the BMN coupling $\lambda'$ fixed:
\bea\label{lambdaprime}
\lambda'=\frac{\lambda}{J^2}\, .
\eea
Let us derive the famous BMN spectrum from our novel string Bethe
ansatz \eqref{BAB}, \eqref{SM}. The derivation is identical
to the one for the long range spin chain for gauge theory \cite{BDS}. 
One easily sees that the momenta scale like $p_k \sim 1/J$
while the corresponding phase factors \eqref{pf} scale like
$\varphi(p_k) \sim J$. Inspection of the part of the discrete 
S-matrix \eqref{SM} involving the phase function $\varphi(p_k)$
then shows that {\it there is no scattering of the elementary excitations
in the BMN limit.} In other words, the excitations are too dilute
to feel each other's presence, and behave like independent particles
on a circle. It is an important and non-trivial property of
our conjectured strong coupling string Bethe ansatz that the further
terms, absent in the gauge theory ansatz, in the S-matrix \eqref{SM}, 
involving products of the excitation charges 
$\q_r(p_k)$, also do not lead to
a scattering phase shift in the strict BMN limit, as will be shown
shortly. This explains why in the latter case gauge and string theory
agree to all orders in the  coupling $\lambda'$ \eqref{lambdaprime}.  
Accordingly the Bethe equations \eqref{BAB} simply become in both
string and gauge theory
\bea\label{betheBMN}
e^{i p_k J}=1 \, ,
\eea
and are immediately solved by $p_k=\frac{2 \pi n_k}{J}$, where the
$n_k$ are arbitrary integer mode numbers
satisfying (from momentum conservation) $\sum_{k=1}^M n_k=0$.
The energies $\Delta$ or anomalous dimensions $\Delta_{{\rm g}}$ 
are then found, 
without further work, in, respectively, string and gauge
theory from \eqref{Ener},\eqref{Dim} as the eigenvalues of 
${\bf E}(g)$,${\bf D}(g)$:
\bea\label{engBMN}
\Delta=\Delta_{{\rm g}}=J+\sum_{k=1}^M \sqrt{1+\lambda'\,n_k^2} \, .
\eea

The above argument for the absence of scattering appears to 
be invalid if some of the excitation numbers $n_k=n_j$ are coinciding,
since then $\varphi(p_k) \rightarrow \varphi(p_j)$ in
\eqref{SM}, and therefore the differences 
 $\varphi(p_k) -\varphi(p_j)$ are no longer of order 
${\cal O}(J)$. These cases require special analysis, see \cite{MZ}, and will 
be discussed in the next subsection. For the strict BMN limit one finds
that this subtlety does not invalidate the final result \eqref{engBMN}.

The absence of scattering in the BMN limit shows, in a way, its
comparative triviality. The situation gets much more interesting
once curvature corrections to the pp-wave metric are taken into 
account. These should then correspond to $1/J$ corrections to
the BMN limit. This near-BMN limit has been investigated 
to leading order ${\cal O}(1/J)$ on
the string side in a number of papers, first for two excitations ($M=2$)
\cite{PR},\cite{Call1}, and, very recently,
for three ($M=3$) \cite{Call2}. This required the inclusion of
rather intricate and involved curvature corrections to the 
quantization procedure. Here we will show that our
strong coupling Bethe ansatz is capable of reproducing, for the
$\su(2)$ subsector, all these known results in a few lines of 
calculation\footnote{The present ansatz is not the only one that is capable of
reproducing the near BMN results of \cite{Call1},\cite{Call2}.
Another ansatz that also works was found by N.~Beisert (unpublished).
However, it appears to be very difficult to find a second ansatz
that also reproduces the $\lambda^{\frac{1}{4}}$ strong coupling
behavior of section 4.}.
What is more, we are able to solve explicitly and
without further work the generic $M$-excitation problem to
${\cal O}(1/J)$ in generality, confer \eqref{engnearBMN} below.

Let us then compute the leading correction to the large $J$
Bethe equation \eqref{betheBMN}. One expands the excitation
momenta $p_k$
\bea\label{regexp}
p_k=\frac{2 \pi n_k}{J}+\frac{p^{(2)}_k}{J^2}\, ,
\eea
and works out the leading large $J$ behavior of the phases
\bea
\varphi(p_k)=\frac{J}{2 \pi n_k} \sqrt{1+\lambda'\,n_k^2}+
{\cal O}(J^0)
\eea
and the excitation charges
\bea
\q_r(p_k)=\frac{2 \pi n_k}{J} \left[
\frac{4 \pi }{\lambda' J} 
\left(\sqrt{1+\lambda'\,n_k^2}-1\right) \right]^{r-1}
+{\cal O}(J^{-r-1})\, .
\eea
Then one expands the Bethe equations \eqref{BAB},\eqref{SM}
and verifies, after explicitly summing over the
products of charges, that the scattering correction to the
``free'' BMN limit is indeed a ${\cal O}(1/J)$ effect. One may then
read off the momentum shift
\bea\label{p2}
\frac{p^{(2)}_k}{2 \pi}=
-\sum_{\textstyle\atopfrac{j=1}{j\neq k}}^M
\frac{n_k^2\sqrt{1+\lambda'\,n_j^2}+n_j^2\sqrt{1+\lambda'\,n_k^2}}
{n_k-n_j}\,.
\eea
Let us note that the form of \eqref{p2} manifestly leads to
$\sum_{k=1}^M p_k^{(2)}=0$, as required by momentum conservation.
Lastly, one expands the expression for the energy, \eqref{Dim}
with \eqref{Qr},\eqref{q2} to next-to-leading order, using the 
expansion \eqref{regexp}. With the help of \eqref{p2} one obtains the 
final, general energy for $M$ excitations, valid to order ${\cal O}(1/J)$, 
and for non-coinciding mode numbers $n_k$:
\bea\label{engnearBMN}
\Delta=J+\sum_{k=1}^M \sqrt{1+\lambda'\,n_k^2} 
-\frac{\lambda'}{J}  
\sum_{\textstyle\atopfrac{k,j=1}{j\neq k}}^M \frac{n_k}{n_k-n_j} \left(
n_j^2+n_k^2\sqrt{\frac{1+\lambda'\,n_j^2}{1+\lambda'\,n_k^2}} \right).
\eea
One easily verifies that this expression reproduces the 
near-BMN spectrum obtained by direct quantization for two 
($M=2$, {\it cf.}~\cite{Call1}) and for three 
($M=3$, {\it cf.}~\cite{Call2} eq.(4.10) on page 26) excitations in the
$\su(2)$ sector. 

\medskip
The reader might find it interesting to compare this formula
to the one obtained from the asymptotic gauge theory ansatz of
\cite{BDS}. Using exactly the same procedure as above we
find
\bea
\label{abanearBMN}
\Delta_{{\rm g}}=J+\sum_{k=1}^M
\sqrt{1+\lambda'\,n_k^2}&-&\frac{\lambda'}{J}\sum_{k=1}^M \frac{M\, n_k^2}
{\sqrt{1+\lambda'\,n_k^2}} \\
&-&\frac{\lambda'}{J}
\sum_{\textstyle\atopfrac{k,j=1}{j\neq k}}^M \frac{2n^2_k n_j}{n^2_k-n^2_j}
\left(
n_j+n_k\sqrt{\frac{1+\lambda'\,n_j^2}{1+\lambda'\,n_k^2}} \right).
\nonumber
\eea
This generalizes the formula for $M=2$, as first conjectured by N.~Beisert in 
\cite{BKS} (second reference) and subsequently derived in \cite{BDS}, 
to all $M$ (assuming separated mode numbers).
The last sum in this expression could be formally further simplified, using
antisymmetry under $j \leftrightarrow k$, by dropping the 
terms involving $n_k^2 n_j^2/(n_k^2-n_j^2)$. 
We preferred the present form in order to manifestly exhibit the 
absence of poles at mode numbers related by $n_j=-n_k$.
One can check, using momentum conservation, that \eqref{abanearBMN} 
agrees with the string formula (\ref{engnearBMN}) for {\it any} $M$ 
up to two loops. At three loops they disagree for {\it any} $M$,
as is to be expected by now.

\subsection{Confluent Mode Numbers}

Our general ``generic'' result \eqref{engnearBMN} exhibits pole 
singularities if any two mode numbers are identical, and
is therefore clearly nonsensical in this case. In fact, even the
argument for the absence of scattering phase shifts in the
strict BMN limit are flawed in the presence of confluences. As 
already pointed out in \cite{MZ} the apparent inconsistency
is resolved by the appearance of half-integer powers of
$J$ in the expansion \eqref{regexp} of the momenta $p_k$.
Let us denote by $\nu_k$ the multiplicities of momenta $p_{k,m_k}$ with leading
identical mode number $n_k$, where the numbers $m_k=1,\ldots,\nu_k$
label the nearly degenerate momenta. Clearly one has
\bea
\sum_{k=1}^{M'} \nu_k n_k=0  \qquad {\rm and} \qquad
M=\sum_{k=1}^{M'} \nu_k \,,
\eea
where $M'$ is the number of {\it distinct} mode numbers.
The refined expansion of the momenta reads\footnote{
The behavior $J^{-\frac{3}{2}}$ is found by assuming a
general correction 
$ p_{k,m_k}=\frac{2 \pi n_k}{J}+\frac{p^{(1)}_{k,m_k}}{J^a}$ and
matching powers in the expansions of both sides of the
Bethe equations.}
\bea\label{singexp}
p_{k,m_k}=\frac{2 \pi n_k}{J}+\frac{p^{(1)}_{k,m_k}}{J^{\frac{3}{2}}}
+\frac{p^{(2)}_{k,m_k}}{J^2}+\ldots\, ,
\eea
Upon insertion into the Bethe equations \eqref{BAB},\eqref{SM}
one finds a non-linear system of equations, for each degenerate
sector $k$ where $\nu_k>1$ \cite{MZ}:
\bea
p_{k,m_k}^{(1)}=-4 \pi^2 n_k^2\, \sqrt{1+\lambda'\, n_k^2}
\sum_{\textstyle\atopfrac{\ell_k=1}{\ell_k\neq m_k}}^{\nu_k}
\frac{2}{p_{k,m_k}^{(1)}-p_{k,\ell_k}^{(1)}}\, .
\eea
Luckily this Stieltjes problem can be solved exactly \cite{S,SD}.
A nice method uses Baxter's $Q$-operator.
For a recent discussion of a similar (but different)
equation system of this type see \cite{LZ}.
We find the solution 
\bea
\label{pkmk1}
(p^{(1)}_{k,m_k})^2
=-4 \pi^2 n_k^2 \sqrt{1+\lambda'\, n_k^2}\, u_{\nu_k,m_k}^2 \, ,
\eea
where the $u_{\nu_k,m_k}$ are the $\nu_k$ roots of the 
{\it Hermite} polynomials 
\bea
Q_{\nu_k}(u)=2^{\frac{\nu_k}{2}}\,H_{\nu_k}(\frac{u}{\sqrt{2}})=
\prod_{m_k=1}^{\nu_k} (u-u_{\nu_k,m_k})\, ,
\eea
satisfying the differential equation
$Q''(u)-u\,Q'(u)+\nu_k\,Q(u)=0$. We thus see that the degeneracy
of the momenta $p_{k,m_k}$ is lifted by $\nu_k$ distinct, purely
imaginary shifts into the complex plane, proving the suppression 
of scattering in the strict BMN limit. One also checks that
these shifts do not lead to unwanted $J^{-\frac{1}{2}}$ corrections
to the energies. In order to compute the $1/J$ correction to the energy 
in the confluent case, we first find an expansion of
the dimension formula \eqref{Dim} up to the order $1/J$:
\bea\label{Deli}
\Delta=J&+&\sum_{k=1}^{M'} \nu_k \sqrt{1+\lambda'\,n_k^2}\\
\nonumber 
&+&\frac{\lambda'}{J}  \sum_{k=1}^{M'}\sum_{m_k=1}^{\nu_k}
\frac{\left( p_{k,m_k}^{(1)}\right)^2}
{8\pi^2 (1+\lambda'\,n_k^2)^{3/2}}+
\frac{\lambda'}{J}  \sum_{k=1}^{M'}\sum_{m_k=1}^{\nu_k}\frac{n_k\, p_{k,m_k}^{(2)}}
{2\pi \sqrt{1+\lambda'\,n_k^2}}
\eea
The sum over $m_k$ of $\left( p_{k,m_k}^{(1)}\right)^2$ on the second line
of (\ref{Deli}) can be easily computed by 
using (\ref{pkmk1}) and the following formula for the roots of Hermite polynomials
\bea
\sum_{m_k=1}^{\nu_k}\left( u_{\nu_k,m_k}\right)^2 = \nu_k (\nu_k -1)\ .
\eea
Thus, it is sufficient to find the sum over $m_k$ of $p_{k,m_k}^{(2)}$ 
to compute the $1/J$ correction  to the energy. 
Because of the summation over $m_k$ the contribution of momenta 
with the same mode number $n_k$ drops out, and the Bethe equations can be easily
solved with the following result
\bea
\label{pkmk2}
\sum_{m_k=1}^{\nu_k} \frac{p_{k,m_k}^{(2)}}{2\pi} = -
\sum_{\textstyle\atopfrac{j=1}{j\neq k}}^{M'} \frac{\nu_k\nu_j}{n_k-n_j} \left(
n_j^2\sqrt{1+\lambda'\,n_k^2}+n_k^2\sqrt{1+\lambda'\,n_j^2} \right).
\eea
Combining all the expressions together, we finally find
\bea\label{enneaBMN}
\Delta=J+\sum_{k=1}^{M'} \nu_k \sqrt{1+\lambda'\,n_k^2} 
&-&\frac{\lambda'}{J}\sum_{k=1}^{M'} 
\frac{\nu_k (\nu_k - 1)n_k^2}{2(1+\lambda'\,n_k^2)} \\
\nonumber
&-&\frac{\lambda'}{J}  
\sum_{\textstyle\atopfrac{k,j=1}{j\neq k}}^{M'} \frac{\nu_k \nu_j n_k}{n_k-n_j} \left(
n_j^2+n_k^2\sqrt{\frac{1+\lambda'\,n_j^2}{1+\lambda'\,n_k^2}} \right).
\eea
In the simplest case where $M=3$, with $\nu_1=2$, $\nu_2=1$ and
$n_1:=n$, $n_2=-2 n$ we obtain using (\ref{enneaBMN})
\bea
\nonumber
\Delta&=&J+2 \sqrt{1+\lambda'\,n^2}+\sqrt{1+\lambda'\,4 n^2} \\
\label{threeimp}
& & -\frac{\lambda'\,n^2}{J}\left[
\frac{6+8\lambda'\,n^2}{\sqrt{1+\lambda'\,n^2}\sqrt{1+\lambda'\,4 n^2}}+
\frac{5+4 \lambda'\,n^2}{1+\lambda'\,n^2}
\right].
\eea
It agrees precisely with the result of direct quantization 
in \cite{Call2}, eq.(4.21) on page 30. 

\medskip
For the reader's convenience we also obtained the analogous formula
following from the asymptotic gauge theory ansatz of
\cite{BDS}:
\bea
\Delta_{{\rm g}}=J&+&\sum_{k=1}^{M'} \nu_k \sqrt{1+\lambda'\,n_k^2} 
-\frac{\lambda'}{J}\sum_{k=1}^{M'} 
\frac{\nu_k (\nu_k - 1)n_k^2}{2(1+\lambda'\,n_k^2)} \\
\nonumber
&-&
\frac{\lambda'}{J}\sum_{k=1}^{M'} \frac{M\nu_k n_k^2}{\sqrt{1+\lambda'\,n_k^2}} 
-\frac{\lambda'}{J}  
\sum_{\textstyle\atopfrac{k,j=1}{j\neq k}}^{M'} \frac{2\nu_k \nu_j n_k^2 n_j}
{n_k^2-n_j^2} \left(
n_j+n_k\sqrt{\frac{1+\lambda'\,n_j^2}{1+\lambda'\,n_k^2}} \right).
\eea
One easily checks that it also agrees with the 
string formula (\ref{threeimp}) up to, but not beyond, two loops.
One can also see that at one loop, 
our formula coincides with the one obtained in \cite{MZ}.
In the simplest three-impurity case we obtain using the formula
\bea
\Delta_{{\rm g}}&=&J+2 \sqrt{1+\lambda'\,n^2}+\sqrt{1+\lambda'\,4 n^2} \\
\nonumber
& &
-\frac{\lambda'\,n^2}{J}\left[
\frac{1}{1 + \lambda'\,n^2} + \frac{6}{{\sqrt{1 + \lambda'\,n^2}}} +
\frac{12}{{\sqrt{1 + \lambda'\,4 n^2}}} -
  \frac{8}{{\sqrt{1 + \lambda'\,n^2}}\,{\sqrt{1 + \lambda'\,4 n^2}}}
\right].
\eea

\section{Strong Coupling Limit}
The quantum Bethe equations allow us to analyze the strong
coupling limit $\lambda\to\infty$ with $L\ll\sqrt[4]{\lambda}$. In
this case we should expect to find the famous $\sqrt[4]{\lambda}$
behavior of operators dual to massive string modes. 

We start by considering
the simplest case of two excitations: $M=2$. 
Assuming $p>0$ in the large $g$ limit the charge densities
turn into \bea \q_r(p)\to
g^{-r+1}\frac{2\sin(\frac{1}{2}(r-1)p)}{(r-1)}
\Big(\sqrt{2}\Big)^{r-1} \eea and the sum over $n$ in
the scattering matrix
(\ref{SM}) can be taken explicitly. 
Then eq.(\ref{BAB})
acquires the form 
\bea  \exp\Big(iLp-8i\,{\sqrt{2}}\,g\,\cos (\half
p )\,\log (\cos (\half p)) \Big) =1 \, ,
\label{BAIK} 
\eea 
where we have taken into account that the phase function
$\v(p)\to\infty$ in the large $g$ limit. This equation can be
easily solved assuming that the momentum $p$ has an expansion
\bea
p = \frac{p_0}{\sqrt{g}} + \frac{p_1}{g} + ...
\label{scs}
\eea
Substituting the expansion into (\ref{BAIK}), we get the leading
contribution
$$
\exp\left( i \sqrt{2}p_0^2\right) = 1
$$
and, therefore,
$$
p_0 = 2^{\frac{1}{4}}\,{\sqrt{n\pi}}.
$$
We want to emphasize that $p_0$ is completely determined by the
exponential term in the scattering matrix (\ref{SM}). Note in particular
that the term $Lp\sim \frac{Lp_0}{\sqrt{g}}$ in the l.h.s. of (\ref{BAIK})
does not contribute into the leading asymptotics due to our 
restriction on $L$: $L\ll \sqrt{g}$.

The conformal
dimension $\Delta$ of the operator is obtained from eq.(\ref{Dim}), and
its leading large $\lambda$ asymptotics is given by
\bea
\label{fr}
\Delta =2\left( n^2\lambda\right)^{\frac{1}{4}}\ .
\eea
This is exactly what one expects to find for dimensions of operators 
dual to massive string modes
at level $n$ with masses $m^2 = 4n\sqrt{\lambda}$ \cite{GKP1}.

\medskip

Let us now understand what happens for a generic case of $M$ excitations.
To this end we have to consider the sum 
\bea
\label{sum}
\chi(p_k,p_j)=\sum_{r=0}^{\infty}\Big(\frac{g^2}
{2}\Big)^{r+2}\Big(\q_{r+3}(p_k)\q_{r+2}(p_j)-\q_{r+2}(p_k)\q_{r+3}(p_j)\Big)\, .
\eea
The roots $p_k$ (generically complex)
obey the conservation law $\sum_{k=1}^M p_k=0$ and 
can be grouped into two sets: $p_k^+$ with $\mbox{Re}\, p_k>0$, $k=1,\ldots, m$ 
and $p_k^-$
with  $\mbox{Re}\, p_k<0$, $k=m+1,\ldots, M$. Once again
we expect $p_k$ to scale as in (\ref{scs}). 
It is therefore convenient to first rescale 
$p\to \frac{1}{\sqrt g}p$ and then consider the 
limit\footnote{Concerning eq.(\ref{sum}), 
one should first perform the sum and only then take the
limit with rescaled momenta substituted.} $g\to \infty$.
Note that
the definition
of the charges $\q_r(p_k)$ requires some care and the 
correct prescription for picking up the sign is 
$$
\frac{\sqrt{\sin^2(\half p)}}{\sin(\half p)}=\mbox{sign~Re}\, p \, .
$$ 
Now taking the strong coupling limit we find that the 
scaling dimension is expressed via rescaled momenta as 
\bea
\Delta= \Big(\frac{\lambda}{2\pi^2}\Big)^{\frac{1}{4}}\left(\sum_{k=1}^m 
p_k^+ -\sum_{k=m+1}^{M}p_k^-\right)\, ,
\label{Dimstrong}
\eea
while the Bethe ansatz equations (\ref{BAI}) turn into 
\bea
\label{stBAI}
\sum_{j=1}^m \chi_{kj}^{++}+\sum_{j=m+1}^M \chi_{kj}^{+-}=-\pi n_k\, ,
\eea
where $n_k\geq 0$ are mode numbers. Here we have also introduced the 
concise notation $\chi_{kj}^{\pm\pm}=\chi(p_k^{\pm},p_j^{\pm})$ to keep 
track of the signs. There are also equations involving 
$\chi_{kj}^{--}$ and $\chi_{kj}^{-+}$ but they are of no relevance to us.

Explicit computation of the leading contribution of $\chi_{kj}^{+-}$ gives
\bea
\label{S}
\chi_{kj}^{+-}=\frac{1}{\sqrt{2}}p_k^+p_j^- \, ,
\eea
while the formula for $\chi_{kj}^{++}$ appears to be rather involved. 
Fortunately, as we show below, we do not need it here.

Let us sum eqs.(\ref{stBAI}) over $k$:
$$
\sum_{k=1}^{m}\sum_{j=1}^m\chi_{kj}^{++}+\sum_{k=1}^{m}
\sum_{j=m+1}^M \chi_{kj}^{+-}=-\pi\sum_{k=1}^{m} n_k\, .
$$
Since $\chi_{kj}^{++}=-\chi_{jk}^{++}$ the first sum vanishes and we are left with 
$$
\sum_{k=1}^{m}\sum_{j=m+1}^M \chi_{kj}^{+-}=\frac{1}{\sqrt{2}}
\Big(\sum_{k=1}^m p_k^+\Big)\Big(\sum_{j=m+1}^M p_j^-\Big)=
-\frac{1}{\sqrt{2}}
\Big(\sum_{k=1}^m p_k^+\Big)^2=-\pi\sum_{k=1}^{m} n_k\, ,
$$
where we have used eq.(\ref{S}) and the momentum conservation law. Therefore, we get 
\bea
\label{LM}
\sum_{k=1}^m p_k^+=-\sum_{k=m+1}^M p_k^-=2^{\frac{1}{4}}\sqrt{\pi \sum_{k=1}^m n_k}\, .
\eea
With this formula at hand the scaling dimension eq.(\ref{Dimstrong}) now reads
\bea
\Delta=2\left(\Big( \sum_{k=1}^m n_k \Big)^2 \lambda \right)^{\frac{1}{4}}\, .
\eea
This formula generalizes eq.(\ref{fr}) to the case of arbitrary number of 
elementary excitations and shows that the corresponding ``gauge theory'' 
operators are dual to string modes with masses: $m^2=4n\sqrt{\lambda}$, where 
the level $n$ is now determined by the mode numbers 
of roots with a positive real part: $n= \sum_{k=1}^m n_k$.
One can think of excitations $p_k^+$ and $p_k^-$
as representing right- and left-moving string modes, and eq.(\ref{LM})
as the level matching condition.

The fact that we do not get any restrictions on $n_k$ means that
the quantum Bethe ansatz equations may only describe strong coupling limit of
long operators with large $L$. The simplest way to see that is to notice that
the Konishi operator is dual to the lightest massive string mode
and, therefore, its anomalous dimension should have $n=1$ in eq.(\ref{fr}).
Since for $L=4$ and $M=2$ the only unprotected operator is a Konishi descendant,
quantum Bethe equations valid for all values of $L$ would restrict the level $n$ to
be 1.\footnote{We thank N.~Beisert for an important discussion on this point.}

\section{Wrapping Speculations}
As we have already pointed out the string Bethe ansatz (\ref{BAB}) differs
from the all-loop asymptotic Bethe ansatz of \cite{BDS} by the 
exponential term in the S-matrix (\ref{SM}). This term captures the 
essential dynamics of quantum strings in the large $g,L$ limit.
Note that we still need to assume $L \gg 1$, but we do not (unlike
in the BMN and spinning strings limits) require $L \sim g$.
However, hypothetically assuming the string ansatz to be valid for all 
values of $g$ and $L$ as is, it leads to disagreement with perturbative gauge 
theory. In particular, the known three-loop scaling dimension of the 
Konishi operator \cite{BKS} is not reproduced by the string Bethe ansatz.
In fact, the ansatz looks rather different from all the standard
Bethe ans\"atze for quantum spin chains. 

It was proposed in \cite{SS,BDS} that disagreement between 
gauge and string theory predictions might be due to neglecting
the wrapping interactions in gauge theory.
At the $k$-th order of perturbation theory the dilatation operator 
is given by the
sum of multi-spin interactions, each of them involving up to $k+1$ neighboring
sites of the spin chain. Therefore, with increasing $k$ the non-locality
of the interactions grows until it encompasses,
at $k=L-1$, all lattice sites of the chain.
Starting at this loop order the interactions wrap around the chain 
and the assumptions of the asymptotic Bethe ansatz break down. 
Since the string predictions are derived
assuming that both $g$ and $L$ are large, the comparison with gauge theory
requires resummation of the gauge theoretic perturbative expansion, $i.e.$
inclusion of the wrapping interactions \cite{SS,BDS}.
At present their structure is unknown. Moreover, the very notion of 
integrability
becomes more subtle because the locality of the spin chain model is lost.
In this situation one can try to get some insight on the possible structure
of the full, nonasymptotic gauge Bethe ansatz from our string theory
ansatz.

Looking at the S-matrix (\ref{SM}) we see that the only explicit dependence
on the coupling constant $g$ appears in the sum entering the exponential terms.
All other dependence is hidden in the phase function and charges. The simplest 
guess for modifying the string scattering matrix so that it becomes compatible
with gauge theory seems to consist in replacing the coefficients 
$\Big(\frac{g^2}{2}\Big)^{r+2}$ by more general functions $c_r(g,L)$ 
which depend on both $g$ and $L$. Thus, we are 
led to consider the Bethe ansatz (\ref{BAB}) with the following S-matrix
\bea
\label{SMI}
S(p_k,p_j)&=&
 \frac{\varphi(p_k)-\varphi(p_j)+i}{\varphi(p_k)-\varphi(p_j)-i}\times \\
&\times &
\exp\Big(2i\sum_{r=0}^{\infty}c_r(g,L)
\big(\q_{r+2}(p_k)\q_{r+3}(p_j)-\q_{r+3}(p_k)\q_{r+2}(p_j)\big)
\Big)
\,
 .
\nonumber
\eea
Once again we assume that the phase function $\v(p)$, the excitation
charges $\q_n(p)$ and the total charges
$\Q_n$ are literally the same as appeared in the
asymptotic Bethe ansatz, and given by eqs.(\ref{pf}), (\ref{qr})
and (\ref{Qr}). The exponential term in the S-matrix depends now 
on an infinite set of functions $c_r(g,L)$. From the gauge theory 
point of view the role of this term might hopefully account for
the wrapping interactions, and, in the following we will conjecturally
call it the  ``wrapping term''. 

We do not have much to say about the explicit form of 
the functions $c_r$; however, in order to be consistent with both
perturbative gauge theory and the known string theory results
these functions have to exhibit the following properties:\footnote{
It is easy to find an example of a function with such properties. For instance
one can take (cf. \cite{BDS})
\bea
\nonumber
c_r(g,L)= \left(\frac{g^2}{2}\right)^{r+2}\tanh^{2(L-r-3)}(g) \,
\quad \mbox{or} \quad  c_r(g,L)= \left(1+\frac{g^2}{2}\right)^{r+2}
\frac{g^{2(L-1)}}{(1+g^2)^{(L-1)}}\, .
\eea
}
\begin{itemize}
\item $c_r(g,L)\to 0$ in the asymptotic limit, $L\to \infty$ and $g$ is held finite;
\item $c_r(g,L)\sim {\cal O}(g^{2(L-1)})$ 
in the perturbative gauge theory, $g\ll 1$ and $L$
is finite;
\item $c_r(g,L)\to \left(\frac{g^2}{2}\right)^{r+2}$ in the limit
$L,g\to \infty$ and $\frac{g}{L}$ is held finite;
\item $c_r(g,L)\to \left(\frac{g^2}{2}\right)^{r+2}$ in the strong coupling limit,
$g\to \infty$ and $1\ll L\ll \sqrt{g}$.
\end{itemize}
Let us now motivate these asymptotic properties of $c_r$.
First of all, we demand the vanishing of the $c_r$ in the asymptotic limit 
to guarantee
that the S-matrix (\ref{SMI}) reduces to 
the S-matrix of the gauge theory asymptotic Bethe ansatz.
The second property of $c_r$ ensures that 
the wrapping interactions do not show up at least
up to the $L$-th order of perturbation theory, which is another 
important feature of gauge theory. 
The third and the fourth properties are needed to recover our string Bethe 
ansatz at strong coupling and, in particular, to obtain the $\sqrt[4]{\lambda}$
asymptotics of scaling dimensions. Thus, the functions $c_r$ might allow to
interpolate smoothly between gauge and string theories, and, for this reason,
it is natural to term the equations (\ref{BAB}),(\ref{SMI}) 
``interpolating Bethe ansatz''.

One could hope that the explicit form of the functions $c_r$
might be determined by the exact quantization of strings 
on $AdS_5\times S^5$. Moreover, 
the Bethe equations themselves provide a very tight restriction on a possible 
quantization ansatz.  
On the other hand, we expect that the interpolating Bethe ansatz
might correctly account for wrapping interactions in gauge theory,
and, therefore, encode the gauge theory perturbative spectrum. 
With wrapping interactions turned off the equations
(\ref{BAB}),(\ref{BAI}) reduce to the ones  
of the asymptotic Bethe ansatz and, therefore, immediately reproduce 
all known results in perturbative gauge theory. Thus, the true test of 
the validity of the interpolating Bethe ansatz in gauge theory 
essentially relies on the currently unknown structure of the
wrapping interactions.  

The conjecture of the interpolating Bethe ansatz implies, quite remarkably,
that from the gauge theory point of view, the $\sqrt[4]{\lambda}$ 
asymptotics is entirely due to the wrapping term. For the near-BMN
limit it leads to an important modification of the result obtained
by the asymptotic Bethe ansatz.

To conclude this section let us note that 
our proposal of the interpolating Bethe ansatz is just a first step towards
an exact quantization of strings on $AdS_5\times S^5$ in the 
$\su(2)$ subsector.
It is important to find internal consistency conditions for 
eqs.~\eqref{BAB},\eqref{SM}.
This might help to understand if other integrable systems, 
in particular spin chain models,
can be described by similar equations. 
In this respect the Inozemtsev long-range spin chain
seems to be a natural candidate for investigating this point.

\section{Summary and Conclusions}
In this paper we proposed a novel set of Bethe ansatz equations.
These equations can be thought of as encoding the quantum spectrum of 
superstring theory on $AdS_5\times S^5$ at large tension and
restricted to the large charge states from the closed $\su(2)$ subsector.
Important evidence for such an interpretation comes from our 
study of the thermodynamic, near-BMN, and strong coupling limits. 

\medskip 
In the thermodynamic limit (\ref{BAB}),(\ref{SM}) reproduce the 
energies of classical spinning strings, and in the near-BMN limit they 
give the $1/J$ correction to the BMN energy formula which agrees and extends 
the results obtained by quantizing string theory in the near plane-wave limit. 
Finally, in the strong coupling limit they reproduce the correct $\sqrt[4]{\lambda}$
behavior of anomalous dimensions. Thus, these equations are compatible 
with our current knowledge of the string theory spectrum on
$AdS_5\times S^5$. 
It would be interesting to find proper generalizations 
of our $\su(2)$ Bethe equations to other closed subsectors. 

\medskip 
We also suggested a ``minimal'' deformation of our string Bethe equations
which might hopefully describe a smooth interpolation between perturbative 
gauge theory and string theory at large tension. These interpolating Bethe
equations involve an infinite set of functions $c_r(g,L)$ known only
in certain asymptotic regimes. We speculate that these functions might 
capture the dynamics of gauge theory wrapping interactions and lead to the 
final agreement of gauge and string predictions.

\medskip
Another, more disappointing scenario would be that free strings on
$AdS_5\times S^5$ and planar ${\cal N}=4$ Super Yang-Mills theory
are described by similar, but {\it different} integrable systems.
If true, this would disprove the AdS/CFT conjecture \cite{M}.

\medskip 
There are several immediate questions one can ask. Here we computed the 
$1/J$ correction to the BMN energy formula for the general case of $M$ impurities. 
The results of exact quantization around a
plane-wave background are known only 
for the states with two and three impurities \cite{Call1,Call2}.
It would be interesting to generalize the calculation of \cite{Call1}
to the case of four (and more)
impurities to check the predictions of the string Bethe ansatz.  

\medskip
Note also that the functions $c_r$ governing the 
interpolating Bethe ansatz can be partially fixed from string theory by 
analyzing $1/J$  and $1/\sqrt{\lambda}$ corrections.
There are several ways to proceed. 
One is to determine $1/J^2$ correction to the BMN
energy formula by using the techniques developed in \cite{Call1}. 
A step in this direction has been recently made in \cite{Swanson}, although 
many details still have to be clarified. Another way is to compute 
the $1/J$ correction to the energy of the two-spin folded string 
along the lines of \cite{FT}. One can also develop the quantization 
of the light-cone Hamiltonian of string theory on $AdS_5\times S^5$ 
\cite{MTT} to compute $1/\sqrt{\lambda}$ corrections 
to masses of string states in the large tension limit. 
All these computations would provide 
further stringent tests of our interpolating Bethe ansatz conjecture. 

\medskip
Of course, one of the most ambitious problems is to {\it derive}
integrability along with the corresponding complete quantum Bethe 
ansatz by exactly quantizing strings on $AdS_5\times S^5$.

\section*{Acknowledgments}
We are very grateful to Niklas~Beisert and Arkady~Tseytlin 
for many useful discussions and helpful comments. 
S.~F. wishes to thank the {\it Max-Planck-Institut
f\"ur Gravitationsphysik} for the warm hospitality.
The work of  G.~A.~was supported in
part by the European Commission RTN programme HPRN-CT-2000-00131
and by RFBI grant N02-01-00695. The work of S.~F.~was supported in part
by the 2004 Crouse Award.


\end{document}